\documentclass[aps,prb,reprint,superscriptaddress]{revtex4-1}
\usepackage{amsfonts}
\usepackage{psfrag}
\usepackage{amsmath}
\usepackage{graphicx}
\usepackage{verbatim}

\begin{document}

\newcommand{\corund}{$\alpha$-Al$_2$O$_3$}
\newcommand{\beryl}{Be$_3$Al$_2$Si$_6$O$_{18}$}
\title{Experimental evidence of thermal fluctuations 
on the X-ray absorption near-edge structure at the aluminum \textit{K}-edge}

\author{D. Manuel}
\email{damien.manuel@impmc.upmc.fr}
\author{D. Cabaret}
\author{Ch. Brouder}
\author{Ph. Sainctavit}

\affiliation{Universit\'e Pierre et Marie Curie (UPMC), 
IMPMC, UMR CNRS 7590, 4 place Jussieu, 75252 Paris Cedex 05, France}

\author{A. Bordage}
\affiliation{Wigner Research Centre for Physics, 
Hungarian Academy of Sciences H-1525 Budapest, P.O.B. 49., Hungary}

\author{N. Trcera}
\affiliation{Synchrotron SOLEIL, BP 48, 91192 Gif sur Yvette, France}

\date{\today}

\begin{abstract}
After a review of temperature-dependent experimental 
x-ray absorption near-edge structure (XANES) and related
theoretical developments,
we present the Al $K$-edge XANES spectra of corundum and beryl for temperature 
ranging from 300K to 930K. These experimental results provide a first evidence
of the role of thermal fluctuation in XANES at the Al $K$-edge especially
in the pre-edge region.
The study is carried out by polarized XANES measurements of single
crystals.
For any orientation of the sample with respect to the x-ray beam, 
the pre-edge peak grows and shifts to lower energy with temperature. 
In addition temperature induces modifications in the position and intensities of the main
XANES features.
First-principles DFT calculations are performed for both compounds. They show that the
pre-edge peak originates from forbidden 1$s$$\rightarrow$$3s$
transitions induced by vibrations.
Three existing theoretical models are used to take vibrations into
account in the absorption cross section calculations:
i) an average of the XANES spectra over the thermal displacements of the 
absorbing atom around its equilibrium position, 
ii) a method based on the crude Born-Oppenheimer approximation 
where only the initial
state is averaged over thermal displacements, 
iii) a convolution of the spectra obtained for the
atoms at the equilibrium positions with an approximate phonon spectral function.
The theoretical spectra so obtained permit to qualitatively understand the origin
of the spectral modifications induced by temperature. However the correct treatment
of thermal fluctuation in XANES spectroscopy requires more sophisticated theoretical tools.
\end{abstract}


\maketitle

\section{Introduction and state of the art}

X ray absorption near-edge structure (XANES) spectroscopy 
is a powerful technique to probe the empty states in solids and
 to determine the local structure around a selected atom.
The interpretation of $K$-edge XANES
spectra is not straightforward and often requires the use of simulation tools, 
which are traditionally based on the density functional theory (DFT). 
XANES calculations for inorganic solids usually consider
the atoms at fixed positions, even if in reality atoms are subjected 
to quantum thermal fluctuations that reduce to the zero point motion 
at $T$=0K. 
Recently, it has been theoretically
shown that vibrations could have a spectacular effect at the Al $K$-edge,
allowing 1$s$ to 3$s$ dipole transitions.\cite{Brouder10}
Such forbidden transitions were previously invoked to be responsible
for the pre-edge feature that occurs at the Al $K$-edge in several
alumino-silicate minerals\cite{Li95} and at the Si $K$-edge in  
silicon diphosphate.\cite{Li94}
Moreover, Ankudinov and Rehr stated that
atomic displacements of order of typical
Debye-Waller factors could reveal some forbidden transitions.\cite{Ankudinov05}
Consequently, the fact that vibrations may be able to induce additional peaks
(in the pre-edge or at higher energies)
suggests that XANES could be considered as a relevant
probe of quantum thermal fluctuations. It is already the case in organic molecules,
the vibronic fine structure of which can be observed in XANES or NEXAFS 
(Near-Edge X-ray Absorption Fine Structure) spectra.
For instance in the last ten years, XANES spectroscopy has revealed
the vibronic structure of 
naphtalene,\cite{Minkov04} biphenyl\cite{Minkov05} and even larger molecules
such as NTCDA\cite{Scholl04} at the C $K$-edge, the vibronic structure of
acetonitrile\cite{Carniato05} and acrylonitrile\cite{Ilakovac08} at the N $K$-edge,
and more recently the one of halogenated acenaphthenequinone\cite{Schmidt11} 
at the O $K$-edge. The vibronic component of the lowest core-excited 
state of OH and OD was observed at the O $K$-edge as well.\cite{Stranges02}

In this paper, we focus on the thermal effects in two inorganic crystalline 
solids at the Al $K$-edge: corundum (\corund) and
beryl (\beryl). Polarized XANES spectra have been  measured
on single-crystals
for temperatures ranging from 300K to 930K. Our purpose is to show
that thermal fluctuations induce substantial spectral modifications, 
especially in the pre-edge region. Very few  studies were carried out
in inorganic solids towards the same goal. As far as we know, 
thermal effects measured in XANES spectroscopy mainly concern
the Ti $K$ pre-edge region in oxides. Such a temperature dependence 
was first observed by Durmeyer \textit{et al.}\cite{Durmeyer90} in TiO$_2$ (rutile), 
Li$_{4/3}$Ti$_{5/3}$O$_4$ and LiTi$_2$O$_4$. The main pre-edge peak 
was found to grow with temperature, and to slightly shift to lower energy 
for the first two compounds. Later, the temperature dependence 
of the polarized pre-edge structure at the Ti $K$-edge in rutile 
was measured at low and room temperature.\cite{Durmeyer10}
Thermal effects were essentially visible 
through an increase of intensity of the first two pre-edge peaks. 
The temperature dependence of the electric dipole and 
quadrupole contributions of the Ti $K$-edge in rutile was investigated
in the 6K-698K range
by Collins and Dmitrienko in 2010 at the Diamond Light Source.\cite{Collins10}
Nozawa \textit{et al.}\cite{Nozawa05} measured the temperature dependence 
of the Ti $K$-edge XANES in SrTiO$_3$ from 15K to 300K, 
showing an increase of the first two pre-edge peaks along
with a slight shift to lower energy of the second pre-edge peak and of 
the absorption edge. These effects were attributed to random thermal vibrations 
of Ti.\cite{Nozawa05}

Apart from these last examples at the Ti $K$-edge, 
most of temperature-dependent XANES measurements were carried out 
in order to study phase transition mecanisms or to determine the local
structure below and above the transition temperature. 
For instance in some perovskite crystals, 
significant spectral changes in the $K$ pre-edge region
are induced by temperature, and 
are interpreted in relation to phase transition. 
The literature reports temperature-dependent x-ray absorption spectra 
(including pre-edge) in perovskite titanates at the Ti 
$K$-edge,\cite{Ravel93,Ravel95,Ravel97,Vedrinskii97,Sato05,Hashimoto07}
in perovskite manganites at the Mn $K$-edge\cite{Bridges00,Qian00} and
at the O $K$-edge,\cite{Mannella05,Tsai09} 
in zirconates at the Zr $K$-edge,\cite{Vedrinskii06}
niobiates at the Nb $K$-edge,
\cite{Shuvaeva99,Shuvaeva01,Shuvaeva03,Shuvaeva04,Lemeshko07}
in the La$_{1-x}$Sr$_x$FeO$_{3-\delta}$
system at the Fe $K$-edge,\cite{Deb06} in a La(Fe,Ni)O$_3$ solid oxide fuel cell
cathode at the O $K$-edge,\cite{Braun11} and more recently in the
Pr$_{0.5}$Ca$_{0.5}$CoO$_3$ cobaltite at the Ca $L_{2,3}$-edges and 
the Pr $L_3$ and $M_{4,5}$-edges.\cite{Herrero-Martin11}
Phase transition studies using temperature-dependent XANES are
not restricted to the perovskite structure. In particular 
Cu $K$-edge and La $L$-edges spectra in La$_{2-x}$Sr$_x$CuO$_4$ cuprates 
were recorded at room temperature and at 78K.\cite{Hidaka03} 
An angular dependence XANES study was performed at the V $K$-edge in VO$_2$ 
for temperatures lower and higher than the metal-insulator
transition temperature at 68$^\circ$C.\cite{Poumellec94}
Several studies reported in literature deal with oxide glass and melt structures.
For instance, the local structural environment of Ti in Na-, K-, and Ca-titanosilicate 
glasses and melts was determined by Ti $K$-edge x-ray absorption spectroscopy 
at temperatures ranging from 293-1650 K.\cite{Farges96}
The Zr $K$-edge in a ZrO$_2$-MgO-Al$_2$O$_3$-SiO$_2$ glass
was measured for various temperatures around 
$T_g$ (1085K).\cite{Dargaud10}
The crystalline and melt structures of Al$_2$O$_3$ and MgAl$_2$O$_4$ 
were investigated using temperature dependent XANES
spectroscopy at the Al and Mg $K$-edges.\cite{Neuville09}
A similar study concerns the crystal, glass and melt structures in
the CaO-MgO-Al$_2$O$_3$-SiO$_2$ system at the Al, Si and Ca 
$K$-edges.\cite{Neuville08}
Again at the Si $K$-edge, the structure of SiO$_2$ polymorphs were studied
using XANES spectroscopy from room temperature to 2030K, 
above the melting point at 2000K.\cite{deLigny09}
Furthermore, at the oxygen $K$-edge in liquid water, variations were
observed between the room temperature spectrum and the 90$^\circ$C one 
and found to be related to a change of local environment 
of a significant amount of molecules.\cite{Wernet04}

Temperature dependent x-ray absorption spectroscopy was 
also used to study chemical reactions. 
For instance, in catalysis, the local structure of Al in
zeolites was examined to understand the processes 
that take place during steam activation of these complex cagelike porous compounds.
Hence Al $K$-edge measurements in several zeolites were carried out
in the range 300K-975K, in vacuum or in a flow of helium saturated 
with water,\cite{vanBokhoven03,vanBokhoven05,Omegna05,Agostini10}
and in other working conditions of the catalysis.\cite{Aramburo12}
In Refs.~\onlinecite{vanBokhoven03,vanBokhoven05,Agostini10,Aramburo12} a pre-edge peak
appears when temperature is increased. The presence of this Al $K$ pre-edge peak 
at high temperature is interpreted as the signature of 
three-fold coordinated Al in the zeolite structure. Other temperature effects 
are observable at higher energies in the XANES region, and attributed to the transformation
of a given amount of octahedrally coordinated Al 
to tetrahedrally coordinated Al.\cite{vanBokhoven05,Omegna05}
Recently, in a totally different field,
the temperature dependence of Fe $K$-edge XANES spectra of FePt/Fe$_3$O$_4$ 
nanoparticles has been measured from 300K to 870K in order to understand
the annealing process, which enables the building of
magnetic nanocomposites with combined magnetic properties.\cite{Figueroa11}
A strong temperature dependence was observed in Fe $L_{2,3}$-edges 
spectra of Fe impurities in MgO thin films in the range 77K-500K; it was 
attributed to the thermal population of low-lying
Fe $3d$ excited states, that are present due to the spin-orbit coupling.\cite{Haupricht10}
At last, the valence state of Yb in YbC$_2$, investigated by Yb $L_3$ XANES 
spectroscopy at low and high temperatures, is found to be stable from
15K to 1123K, so providing new information about this long-known compound.\cite{Link11}

In all the references cited in the previous two paragraphs, the role of nuclear motion 
in the temperature-dependent XANES spectra was not investigated.
A thorough study of thermal fluctuations could then bring new insights in the
interpretation of the temperature-dependent spectral features 
whatever the purpose of the experiments is (catalysis, phase transition, etc.).

The modeling of nuclear motion 
in the absorption cross section of solids is a challenging task.
In particular, the presence of phonon modes in crystalline solids
generates a dynamic disorder, which has a completely different behavior
from the static disorder that can be found in glasses. 
Far from the edge, in the extended x-ray absorption fine structure
(EXAFS) region, vibrations are taken into account
through a Debye-Waller factor $\exp{(-2k^2\sigma^2)}$.\cite{Vila07,Vila12}
Since this factor vanishes for $k \simeq 0$, it cannot reproduce well 
the thermal effects in the XANES region and especially in the pre-edge region.
Fujikawa and co-workers showed
in a series of papers of increasing sophistication\cite{Fujikawa96,Fujikawa99,Arai07}
that the treatment of vibrations could be achieved 
by the convolution of the ''phononless'' x-ray absorption spectrum
with the phonon spectral function. To our knowledge, this theoretical work has
not been applied to any real case yet. A basic and simple idea to 
take vibrations into account is to calculate the absorption cross-section
for a configuration where the absorbing atom is shifted
from its equilibrium position.\cite{Ankudinov05,Vedrinskii06} 
For instance, at the B $K$-edge of ABO$_3$ perovskite structures, 
the approach used by Vedrinskii \textit{et al.}\cite{Vedrinskii06} 
to treat the temperature effects is based 
on the assumption that the area of the pre-edge peak is proportional 
to the mean square displacements of the B atom along the O-B-O chain.
In such a way, the symmetry-breaking of the absorbing atom site
generates or increases local $p$-$d$ and $p$-$s$ hybridization.
Brouder and co-workers\cite{Brouder10,Cabaret09} developed
a different approach assuming that vibrational energies are small 
with respect to the instrumental resolution, 
and using the crude Born-Oppenheimer
approximation so that only core-hole motion remains. 
The resulting expression for the absorption cross section 
shows that, at the $K$ edge, vibrations enable 
electric-dipole transitions to 3$s$ and 3$d$ final states, that are not due
to local hybridization with the $p$ states.
In the case of organic molecules, more advanced theories are employed
to reproduce the vibronic structure observed in XANES spectra.
Beyond the Born-Oppenheimer approximation, a vibronic coupling theory
was developed and successfully applied to the C-$1s$ absorption spectra
of ethylene (C$_2$H$_4$)\cite{Koppel97} and ethyne (C$_2$H$_2$)\cite{Kempgens97} 
and isotopomers. 
Besides, many approaches calculate Franck-Condon factors based on 
a vibrational eigenmode analysis in the ground and
excited states. In particular, the linear coupling model 
was used to compute the Franck-Condon factors 
for NEXAFS spectra 
at the O $K$-edge in formaldehyde,\cite{Trofimov01}
at the C $K$-edge in formaldehyde,\cite{Trofimov03}
naphtalene\cite{Minkov04}, 
gaseous cyclopropane,\cite{Duflot06}
and acetic acid,\cite{Duflot09}
at the N $K$-edge in gaseous pyridine,\cite{Kolczewski01}
acetonitrile\cite{Carniato05} and 
acrylonitrile.\cite{Ilakovac08}
In Ref.~\onlinecite{Ilakovac08} a direct calculation
of the Franck-Condon amplitudes is also performed. 
However, the Franck-Condon approximation
ignores the impact of nuclear motion on the electronic
transition amplitude. To first order, this impact is
referred to as the Herzberg-Teller effect.\cite{Herzberg33}
To go beyond the Franck-Condon approximation, 
the nuclear degrees of freedom of molecules was modeled using various 
molecular dynamics (MD) techniques.
In doing so, atomic configurations are generated at finite temperature 
and then used as input in cross-section calculations. Finally, the individual spectra 
associated to each sampled atomic configuration are averaged over.
Using this methodology, vibrations were included 
in XANES calculations at the N $K$-edge of several prototype molecules:
upon classical MD sampling,\cite{Uejio08,Schwartz09,Schwartz10}
and upon path-integral molecular dynamics (PIMD).\cite{Schwartz09,Fatehi10} 
PIMD sampling enables to consider the quantum nature of 
nuclear motion, which has been found to be of noticeable importance 
to accurately simulate XANES of N$_2$,\cite{Schwartz09}, 
and s-triazine and glycine.\cite{Fatehi10}
The sensitivity of photoabsorption spectroscopy 
to a quantum treatment of nuclear motion
was also highlighted in the optical range\cite{DellaSala04} and  
in the UV range.\cite{Kaczmarek09} 
The MD methodology is not restricted to organic molecules;
simulations of the density and temperature dependence 
of XANES in warm dense aluminum plasmas were achieved
using \textit{ab initio}\cite{Mazevet08,Recoules09} and 
classical\cite{Peyrusse10} MD and recently they have been compared 
to experiments.\cite{Benuzzi-Mounaix11,Levy12}

In the present study, DFT calculations 
based on plane-wave formalism
are performed in order to understand
the spectral modifications induced by temperature that are observed experimentally. 
Reference theoretical Al $K$ XANES spectra of corundum and beryl 
are first obtained considering the atoms at their equilibrium positions. 
Then three existing methods are used to account for vibrations: \textit{(i)} 
a calculation considering thermal displacements of the absorbing atom around 
its equilibrium position, 
\textit{(ii)} the method developed by Brouder \textit{et al.},\cite{Brouder10}
\textit{(iii)} a convolution of the spectra obtained for the
atoms at the equilibrium positions with an approximate phonon spectral function.

The paper is organized as follows. Section~\ref{sec:methods} is dedicated to the methods.
First the temperature-dependent XANES experiments at the Al $K$-edge of
corundum and beryl are described,
with a specific attention paid to the self-absorption correction procedure. 
Second, the three theoretical DFT methods used to take the vibrations into account
are detailed. Section~\ref{sec:resexp} is devoted to the description of 
the experimental spectra recorded for temperatures ranging from 300K to 930K. 
In section~\ref{sec:discussion} the experimental results are 
analyzed with the help of the DFT calculations, and discussed in the context of 
the temperature-dependent XANES spectra already reported in the literature.
The conclusion of this work is given in Section~\ref{sec:conclusion}.

\section{Methods}
\label{sec:methods}
\subsection{Experimental setup}

Two single-crystalline samples containing aluminum in 6-fold coordination 
with oxygen are studied. 
The first is a cylindrical synthetic transparent ruby 
($\alpha-$Al$_2$O$_3$ with 18~ppm of Cr and 10~ppm of Ti) 
of 4.0~mm diameter and 1.0~mm thickness, which will thereafter 
be denoted as corundum. 
The second is a parallelepipedic section of a natural green emerald 
from Colombia (Be$_3$Al$_2$Si$_6$O$_{18}$ 
with 238~ppm of Fe, 78~ppm of Cr, 59~ppm of Sc and 36~ppm of Co) 
of 10.0~mm length, 3.6~mm width and 1.2~mm thickness, 
which will be denoted as beryl. 
The samples were analyzed using the CAMPARIS electronic microprobe 
at Universit\'e Pierre et Marie Curie - Paris 6, France.

X-ray absorption experiments were performed at LUCIA beamline 
in the French synchrotron facility SOLEIL.\cite{Lucia06}
The synchrotron was operating in the top-up mode at
2.75~GeV with a current of 400~mA.
The beam spot size was set to 1$\times$2~mm$^2$ and 
the energy range chosen to include the Al \textit{K}-edge 
(1559.6 eV in bulk)\cite{Bearden67} was set from 1550 eV to 1700 eV, 
hence measuring pre-edge, XANES and the beginning of EXAFS. 
Energy selection was performed through the combination 
of an HU52 ``Apple II'' type undulator and a double KTP (011) 
crystal monochromator. 
The pressure in the experimental chamber was 10$^{-5}$ mbar.

The samples were held between a parallelepipedic boron nitride 
furnace and a punched molybdenum lamella fixed to the furnace, 
allowing temperatures ranging from 300K to 950K. 
This holder was fixed on a mobile stage, 
allowing translations along a cartesian $xyz$ coordinate system 
plus a $z$-axis rotation, where the $x$-axis corresponds 
to the beam direction and the $y$-axis to the (horizontal) 
linear polarization $\boldsymbol{\varepsilon}$ of the beam. 
In order to minimize self-absorption effects, 
the sample surfaces should be orthogonal to the beam axis. 
For fluorescence detection, the sample holder has been slightly rotated 
by 15$^\circ$ around $z$-axis.

Corundum and beryl point groups are $\bar{3}m$ and $\frac{6}{m}mm$ 
respectively.\cite{Newnham62,Hazen86} 
This implies that both materials are dichroic in the electric dipole 
approximation.\cite{Brouder90} The single-crystals were set 
on the sample holder in order to measure successively $\sigma_{\parallel}$ 
and $\sigma_{\perp}$ spectra, corresponding respectively 
to $\boldsymbol{\varepsilon}$ parallel and perpendicular to the high symmetry axis, 
i.e. the three-fold symmetry axis for corundum and the six-fold symmetry axis for beryl.

Total fluorescence yield was measured by a four element silicon drift diode 
(SDD) detector with a total active area of 40~mm$^2$ protected 
from infrared and visible radiations by a thin beryllium window. 
In order to maximize the signal/noise ratio, each point was obtained 
after a six second acquisition time and five consecutive spectra were measured 
for each configuration (a configuration consists of a sample, 
an orientation and a temperature). 
This system produces five intensity outputs, one for each of the four SDD 
($I_1$, $I_2$, $I_3$, $I_4$) and one for the incident beam ($I_0$), 
measured before the sample.
For each spectrum, each $I_j$ was then divided by $I_0$, 
followed with a normalization on the 1590-1700 eV energy range 
(spectra are divided by their mean values on this region) 
giving ($I_j$/$I_0$)$_{\mathrm{norm}}$.
Then, ($I_j$/$I_0$)$_{\mathrm{av}}$ was calculated for each configuration 
by averaging five consecutive acquisitions of ($I_j$/$I_0$)$_{\mathrm{norm}}$.
Since self-absorption effects are angle dependent and since the four SDD 
were not at the exact same spatial position, 
the four ($I_j$/$I_0$)$_{\mathrm{av}}$ were not directly comparable.
Self-absorption correction was performed using the formula:\cite{Goulon82,fluo}
\begin{equation}
\sigma_j(\omega) = \frac{\mu_{j} (\omega)}{\mu_{j} (\omega_0)} = 
\frac{N_j(\omega) \left( \beta \frac{\sin{\theta_i}}{\sin{\theta_{f,j}}}  
+ \gamma \right)}{\left( \beta \frac{\sin{\theta_i}}{\sin{\theta_{f,j}}}  
+ \gamma +1 \right) - N_j(\omega)}
\label{eq:fluocorrec}
\end{equation}
in which $\mu_{j}$ is the edge absorption coefficient for detector $j$, 
$N_j(\omega)= \frac{(I_j/I_0)_{\mathrm{av}}(\omega)}{(I_j/I_0)_{\mathrm{av}}(\omega_0)}$ 
is the measured intensity on silicon drift diode $j$, 
normalized at a fixed pulsation $\omega_0$ (chosen after the main edge 
at $\hbar\omega_0$=1653~eV), $\theta_i$=$75^\circ$ is the angle 
between the beam axis and the sample surface, $\theta_{f,j}$ 
is the angle between the sample surface and the outgoing fluorescence beam 
(which depends on each detector $j$ position).
Coefficient $\beta$ is equal to 
$\frac{\mu_{\mathrm{tot}}(\omega_{\mathrm{fluo}})}{\mu(\omega_0)}$ 
where $\omega_{\mathrm{fluo}}$ is the K$\alpha_1$-fluorescence pulsation 
($\hbar \omega_{\mathrm{fluo}}$=1486.7~eV) and $\mu_{\mathrm{tot}}$ 
is the total absorption coefficient, including edge absorption $\mu$ 
and background contribution $\mu_{\mathrm{bg}}$ of other atomic species
in the sample. Finally $\gamma$ is equal to $\frac{\mu_{\mathrm{bg}}}{\mu}(\omega_0)$.
Values for $\beta$ and $\gamma$ are calculated based on crystal stoichiometry, 
atomic masses, unit cell geometry and using experimental tabulated atomic 
values.\cite{Hubbel88} For example, 
$\mu_{\mathrm{tot}}(\omega)$=$\sum_\mathrm{at} \rho_{\mathrm{at}}
\mu_{\mathrm{at}}^{\mathrm{table}}(\omega)/M_{\mathrm{at}}$ 
with $\rho_{\mathrm{at}}$=$N_{\mathrm{at}} M_{\mathrm{at}} 
\mathcal{N}_A / V_{\mathrm{u.c.}}$ 
where the sum is over all the atoms in the unit cell and 
for each atom $\mu_{\mathrm{at}}^{\mathrm{table}}$ is the 
tabulated value in Barns/atom, $M_{\mathrm{at}}$ is the molar mass 
in g.mol$^{-1}$, $N_{\mathrm{at}}$ is the number of atoms `at' 
in the unit cell, $V_{\mathrm{u.c.}}$ is the unit cell volume 
in cm$^3$ and $\mathcal{N}_A$ is the Avogadro constant.
Such a correction gives comparable values of $\sigma_{j}$ for each detector, 
each one being normalized at the point of energy $\hbar \omega_0$. 
By using such a procedure to perform self-absorption correction,
averaging over the four detectors is possible and gives the 
spectra $\sigma_{\parallel}$ and $\sigma_{\perp}$ shown in section \ref{sec:resexp}.
No smoothing  were applied to the spectra.

\subsection{Theoretical methods}
\label{subsec:methodscalc}

In order to explain the experimental spectral features observed 
when temperature is increased, first-principles calculations were 
performed in the density functional theory framework 
using the Quantum-Espresso suite of codes.\cite{QE} 
The 1$s$ core-hole effects were taken into account within 
a 2$\times$2$\times$2 trigonal supercell for corundum, 
containing 80 atoms,\cite{Newnham62} and 
a 2$\times$1$\times$1 hexagonal supercell for beryl, 
containing 58 atoms.\cite{Hazen86}
Troullier-Martins\cite{Troullier91} norm-conserving pseudopotential were used. 
The pseudopotentials of Al and Si were generated using 
the 3$s$, 3$p$ and 3$d$ orbitals as valence states, 
with 2.00~Bohr cutoff radii and considering the $d$ states as local. 
The pseudopotential of the Al absorbing atom was generated with 
only one 1$s$ electron in the Al electronic configuration.
The oxygen and beryllium pseudopotentials were built 
using the 2$s$ and 2$p$ as valence states (local part $p$) 
with cutoff radii of 1.46~Bohr for O and of 2.00~Bohr for Be.
Self-consistent charge densities were calculated in the 
generalized gradient approximation of Ref.\onlinecite{Perdew96}, 
using the PWscf code, at the $\Gamma$ point, with plane-wave energy cut-off of 80~Ry.
Thereafter, local and partial density of states were calculated 
using L\"owdin projections on a 4$\times$4$\times$4 $k$-point grid 
with a Gaussian broadening parameter of 0.3~eV.
XANES cross section calculations were performed using the XSpectra 
code.\cite{Gougoussis09,Taillefumier02}
The all-electron wave-function of the Al absorbing atom was 
reconstructed using the Projector Augmented Wave method,\cite{Blochl94} 
with augmentation region radii of 2.0~Bohr.
The spectra were computed on 4$\times$4$\times$4 $k$-point grid 
with a broadening parameter $\gamma$=0.6~eV.

Three methods were used to take into account the effects of 
vibrations without having to calculate phonon modes.
First, adapting the idea of Ref.\onlinecite{Nozawa05} on vibrations,
only the absorbing atom was moved in the crystal structure with respect
to its equilibrium position. 
This method will be referred to as 'method 1' in the following section. 
As vibrations are concerned, spectra for several absorbing atom 
displacements $\mathbf{R}$ were calculated and averaged with a 
weighting core displacement distribution function $\rho(\mathbf{R})$.
Within the harmonic approximation, 
this distribution function is linked to the thermal parameter matrix 
$[U_{ij}]$ through 
$\rho(\mathbf{R}) = \exp(-\mathbf{R} \cdot \frac{U^{-1}}{2} \cdot \mathbf{R})$. 
The components of the rank-2 tensor $U$ are given by X-ray or 
neutron scattering measurements. In this approximation 
each spectrum is calculated with a static off-center absorbing atom, 
creating hybridization between Al valence orbitals and 
neighboring atoms orbitals, thus modifying valence states of the crystal.
Vibration modeling comes from averaging spectra over a large number 
of absorbing atom displacements.

\begin{figure}
\centering
\includegraphics[width=0.25\textwidth,scale=0.9]{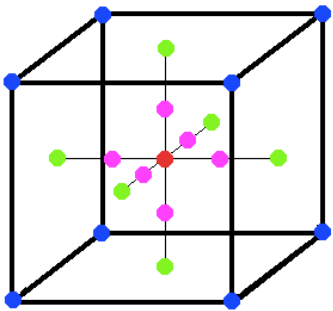}
\caption{\label{fig:cubeintegration} A sketch of the 21-point cubic grid 
used in Eq.~\ref{eq:integral_cube}.}
\end{figure}

The absorbing atom motion was discretized on a cubic 
tridimensional grid consisting of 361 points. 
Equation~\ref{eq:integral_cube} gives the cubic integral approximation
on a grid of 21 points (see Fig.~\ref{fig:cubeintegration}):\cite{Abramowitz64}
\begin{equation}
\begin{array}{rl}
\displaystyle \frac{1}{a^3} \int_{cube} f(\omega,\mathbf{R}) d\mathbf{R} = 
& \displaystyle \frac{1}{360} ( -496 f_m + 128 \sum f_r \\
\displaystyle & \displaystyle + 8 \sum f_f + 5 \sum f_v ) + O(a^6), \\
\end{array}
\label{eq:integral_cube}
\end{equation}
where $a$ is the length of the cube edge, 
$f(\omega,\mathbf{R})$=$\sigma_e(\omega,\mathbf{R}) \times \rho(\mathbf{R})$ 
is the weighted spectrum calculated with absorbing atom displaced 
of $\mathbf{R}$, $f_m = f(\omega,\mathbf{0})$ is the value at the 
center of the cube (red point), $\sum f_r$ is the sum of the values 
of $f$ at the 4 points halfway between the center of the cube and 
the center of each face (pink points), $\sum f_f$ is the sum of 
the values of $f$ at the center of each face (green points) and 
$\sum f_d$ is the sum of the values of $f$ at the vertices of 
the cube (blue points).
The 361 point grid is an augmented 3$\times$3$\times$3 version of 
the 21 point one.
This discretization would lead to a 27$\times$21 point grid, 
which reduces to only 361 points after avoiding double counting 
of similar points.
The value of the integration cube edge $a$ is given
by $2\sqrt{U_\mathrm{eig}}$, where $U_\mathrm{eig}$ are
the eigen values of the $U$ matrix. 
For the 'method 1' calculations, $a$ was chosen equal to 0.3~Bohr radius,
in agreement with the thermal matrix parameters of corundum
and beryl given in Refs.~\onlinecite{Thompson83,Hazen86}.

The second method 
will be referred to as `method 2'.\cite{Brouder10,Cabaret09}
Since vibrational energies are small with respect 
to the instrumental resolution, the closure approximation is made. 
In this approximation, the final vibrational states of the electrons+nuclei system
are summed over. 
Hence atomic motions are assumed not to change the final electronic states
that are calculated at the atomic equilibrium positions. 
Consequently, in `method 2',
only the position of the absorbing atom in the initial state 
is modified by vibrations.
In practice, for a $K$-edge, the $1s$ wave-function of 
the initial state is no more centered on the absorbing atom equilibrium position. 
This shifted $1s$ wave-function is expanded over spherical harmonics 
centered on the equilibrium position. The $\ell$=$1$ component 
allows the transition to $3s$ states.
Contrary to `method 1', this method does not create hybridization in the final state.
Calculating absorption cross section in this method is based on Eq.~(9) 
from Ref.~\onlinecite{Brouder10}. 
The $1s$ wave-function movements are calculated using the same 
core displacement distribution used in `method 1', integrating 
over the same 361 point grid, but using a shorter value for $a$ 
($a$=0.1~Bohr radius as in Ref.~\onlinecite{Brouder10}).

The third method, denoted as `method 3', 
is based on Fujikawa's work.\cite{Fujikawa99} 
Within the Franck-Condon approximation, the effect of vibrations 
on x-ray absorption spectroscopy can be represented as the convolution 
of the phonon spectral function with the x-ray absorption spectrum 
at equilibrium position. Here the phonon spectral function is approximated 
as a Gaussian distribution.
Along with the two precedent methods, convoluted spectra are calculated.
The convolution is performed between the first-principles 
equilibrium spectrum for a configuration and a Gaussian. 
The Gaussian equation is $1/(\sqrt{2\pi}\sigma) \exp(-x^2/(2\sigma)^2)$, 
with parameter $\sigma$ being chosen for each configuration 
in order to reproduce the main edge intensity of method 1 spectra.

Calculated spectra presented in section \ref{sec:discussion} are all 
normalized on the high-energy region, as was done for experimental data.

\section{Experimental Results}
\label{sec:resexp}

\begin{figure*}
\centering
\includegraphics[width=0.49\textwidth]{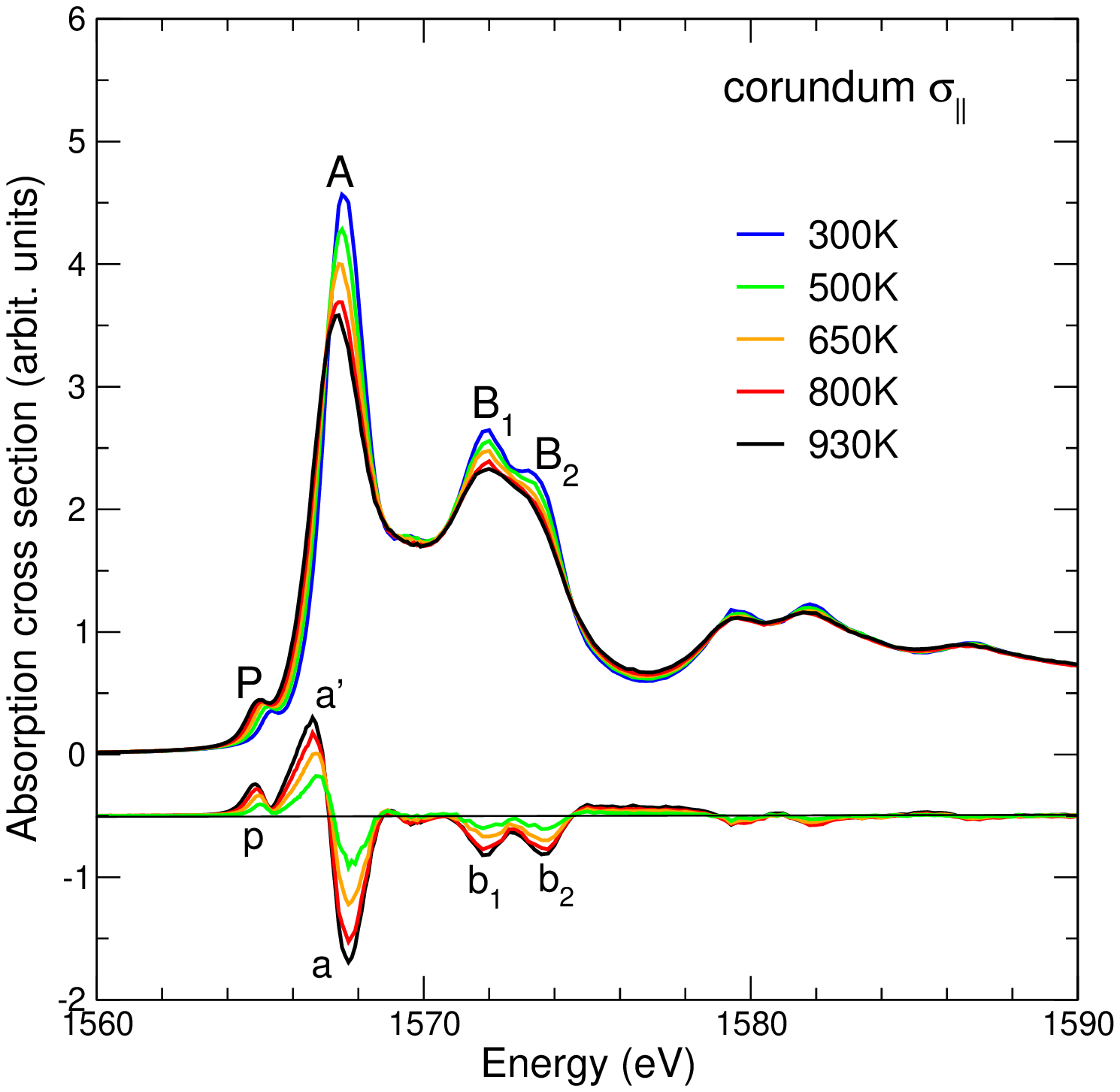}
\includegraphics[width=0.49\textwidth]{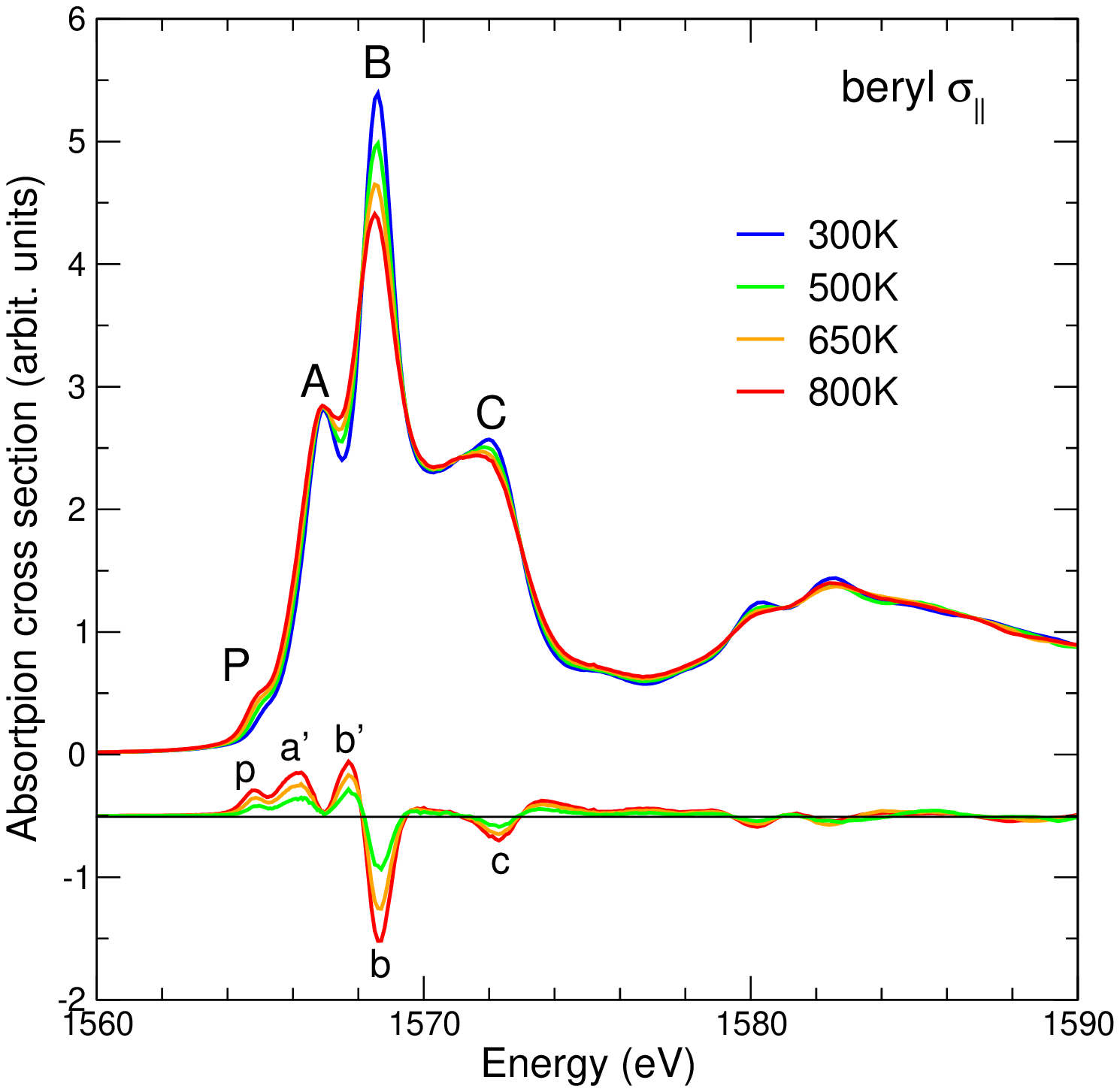}
\includegraphics[width=0.49\textwidth]{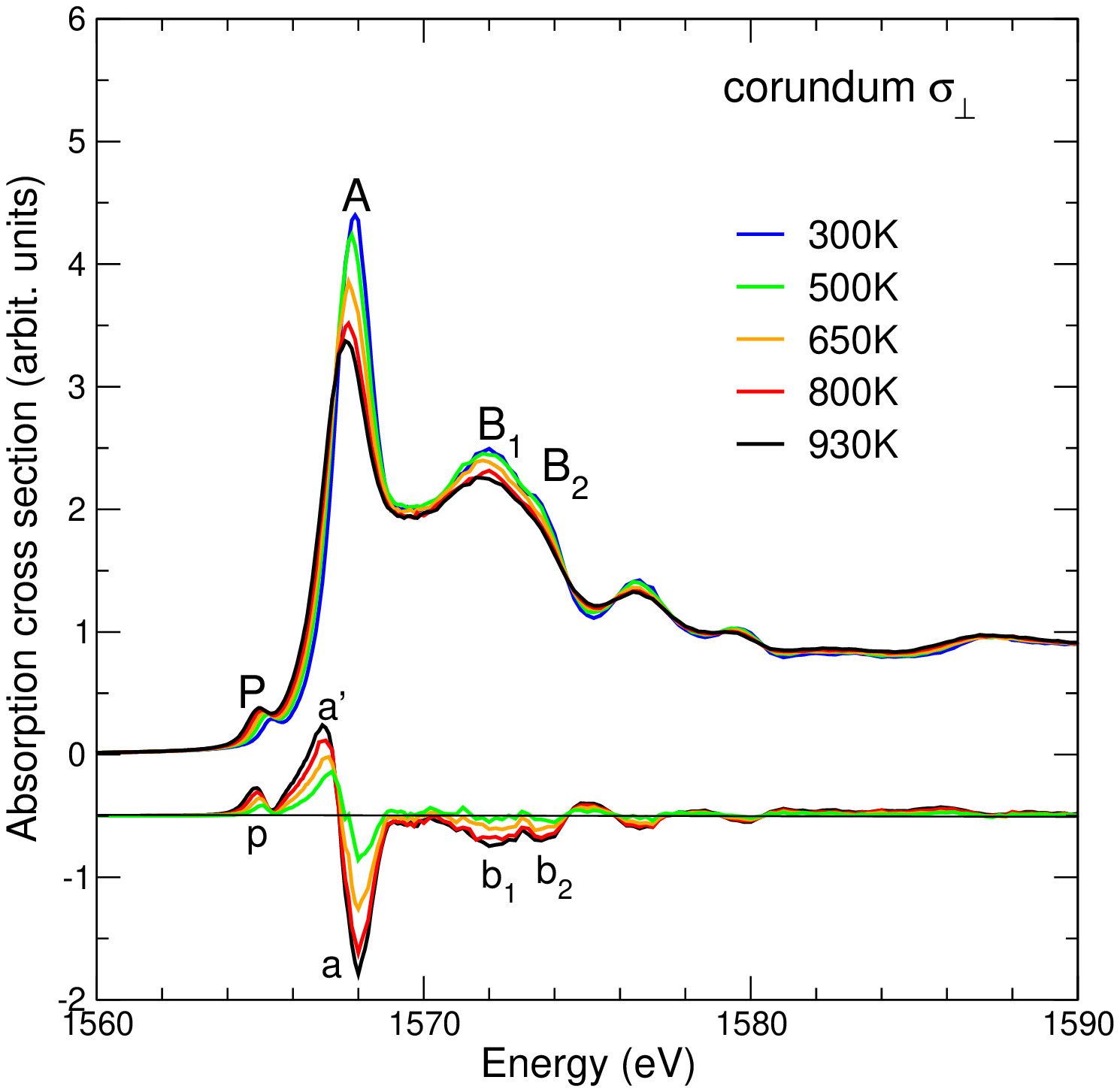}
\includegraphics[width=0.49\textwidth]{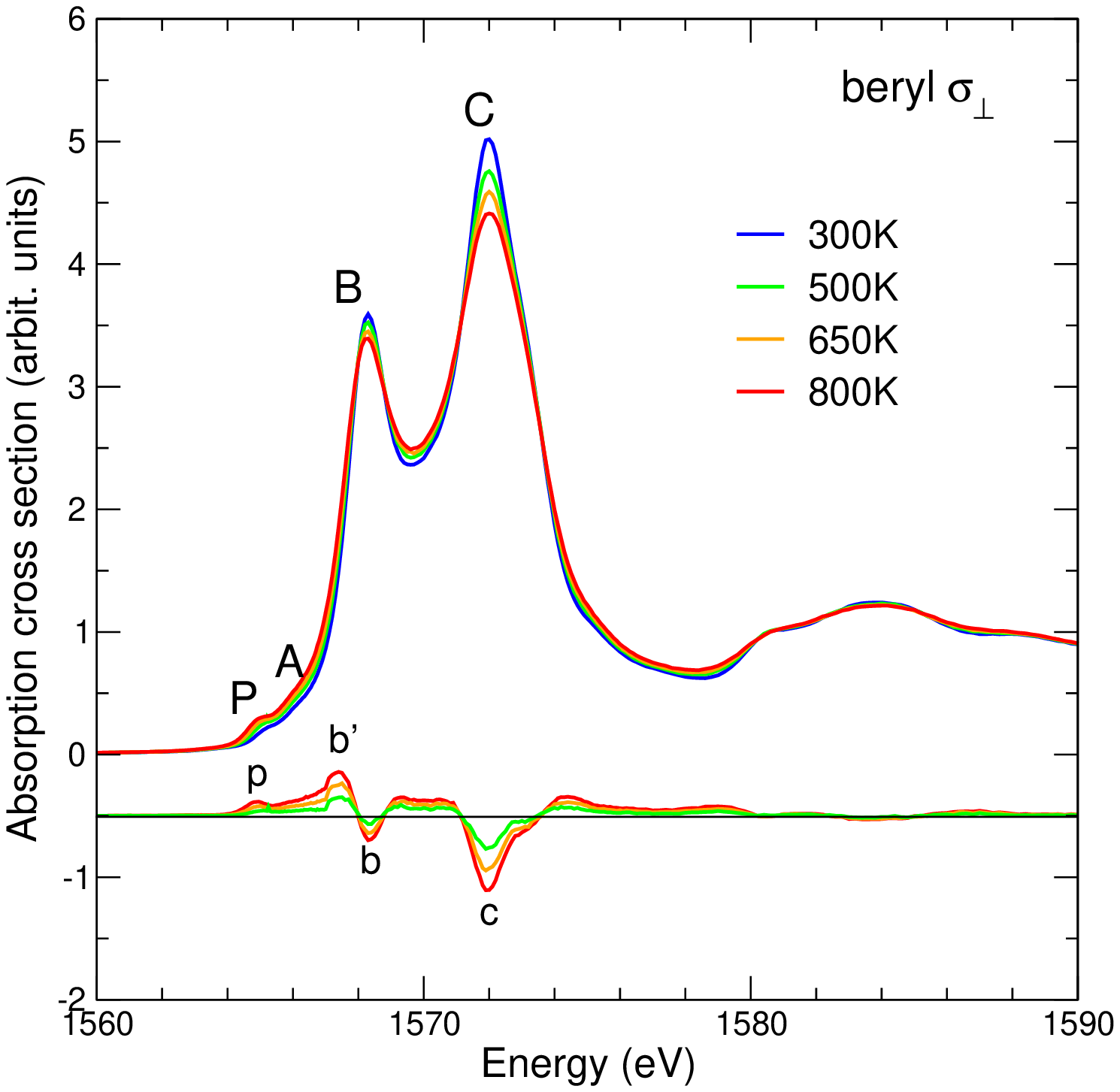}
\caption{\label{fig:resexp}Experimental X-ray absorption polarized spectra
of corundum (left) and beryl (right) at the Al \textit{K}-edge for different
temperatures along with difference of each spectrum with respect to
the 300K reference. The top and bottom panels display the
$\sigma_{\parallel}$ and $\sigma_{\perp}$ spectra, respectively.}
\end{figure*}

Figure \ref{fig:resexp} shows the self-absorption corrected XANES 
spectra of corundum and beryl recorded at different temperatures 
for both orientations. Spectra originating at $y$=$-0.5$ correspond 
to the difference $\sigma(T)$$-\sigma(300\mathrm{K})$.
We observe here for the first time the slow evolution of the Al $K$ 
pre-edge feature with temperature. In addition to the expected increase 
in the peak intensity,\cite{Brouder10,Cabaret09} a shift toward 
lower energy of the pre-edge structure is observed when temperature 
increases. Furthermore, this effect appears for each sample and 
for each orientation and seems to be general,
as well as the decreasing intensities and broadening 
of main peaks with temperature.

Corundum $\sigma_{\parallel}$ and $\sigma_{\perp}$ spectra are 
quite similar up to 1575~eV: they both present a well-resolved 
pre-edge peak \texttt{P}, the main peak \texttt{A} followed by 
the double feature \texttt{B$_1$} and \texttt{B$_2$}. The 300K polarized spectra
are in agreement with those recorded in total electron yield shown 
in Ref.~\onlinecite{Cabaret05}. 
As mentioned before, the intensity of peak \texttt{P} increases with temperature 
while its position is shifted towards lower energy. This is confirmed 
by the peak \texttt{p} in the difference spectra. Peak \texttt{A} also 
varies with temperature: it broadens, its intensity decreases 
and its position is shifted to lower energy (around 0.2~eV 
and 0.3~eV for $\sigma_{\parallel}$ and $\sigma_{\perp}$ respectively). 
The decrease of peak \texttt{A} intensity corresponds to the difference 
spectra peak \texttt{a} and the shift to lower energy to \texttt{a'}. 
Peak \texttt{a'} is also related to the broadening of peak \texttt{A} 
with temperature. Peaks \texttt{B$_1$} and \texttt{B$_2$} 
intensities also diminish and they only form one broad peak at 930K 
in $\sigma_{\parallel}$ and $\sigma_{\perp}$. 

Beryl $\sigma_{\parallel}$ and $\sigma_{\perp}$ spectra present 
much different shapes, unlike the case of corundum. In the energy range 
1560-1575~eV, the $\sigma_{\parallel}$ spectrum exhibits the 
pre-edge peak \texttt{P} and three main features \texttt{A}, 
\texttt{B} and \texttt{C}, feature \texttt{B} being around 
twice more intense than \texttt{A} and \texttt{C} at 300K. 
On the contrary, the $\sigma_{\perp}$ spectrum exhibits 
two main features \texttt{B} and \texttt{C}, preceded 
by a small shoulder \texttt{A} and the pre-edge \texttt{P}.
Peak \texttt{C}  is more intense than peak \texttt{B}.
Peak \texttt{A} and \texttt{B} of 
$\sigma_{\perp}$ are located at 0.7~eV and 0.3~eV lower 
than their corresponding peaks in $\sigma_{\parallel}$.
The pre-edge peak \texttt{P} is less intense in $\sigma_{\perp}$ 
than in $\sigma_{\parallel}$ at 300K and its intensity increases
more in $\sigma_{\parallel}$ than 
in $\sigma_{\perp}$. Indeed the difference peak \texttt{p} 
appears better resolved and more intense in $\sigma_{\parallel}$ 
than in $\sigma_{\perp}$. Peak \texttt{A} of $\sigma_{\parallel}$ 
has a constant intensity with temperature but creates the difference 
peak \texttt{a'} due to broadening.
The weak peak \texttt{A} of $\sigma_{\perp}$ decreases in intensity 
with temperature. The \texttt{B} peak intensity also decreases 
with temperature, as well as a slight shift of 0.1~eV toward 
lower energy, that creates peaks \texttt{b} and \texttt{b'} 
in the difference spectra, more pronounced in $\sigma_{\parallel}$ 
than in $\sigma_{\perp}$. Peak \texttt{C} decreases and broadens
with temperature, leading to feature 
\texttt{c} in the difference spectra.

\begin{figure}[ht]
\centering
\includegraphics*[width=0.40\textwidth]{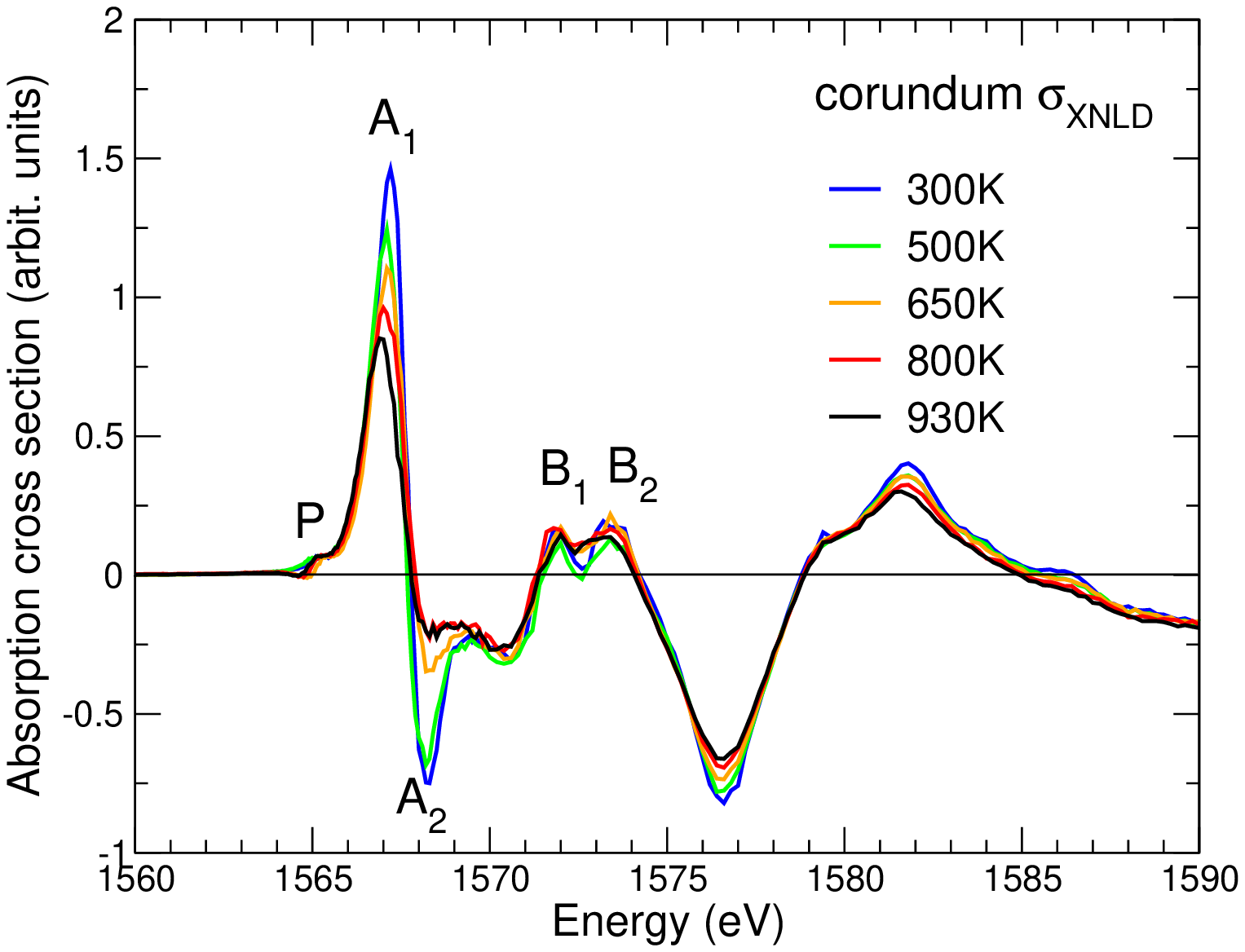}
\includegraphics*[width=0.40\textwidth]{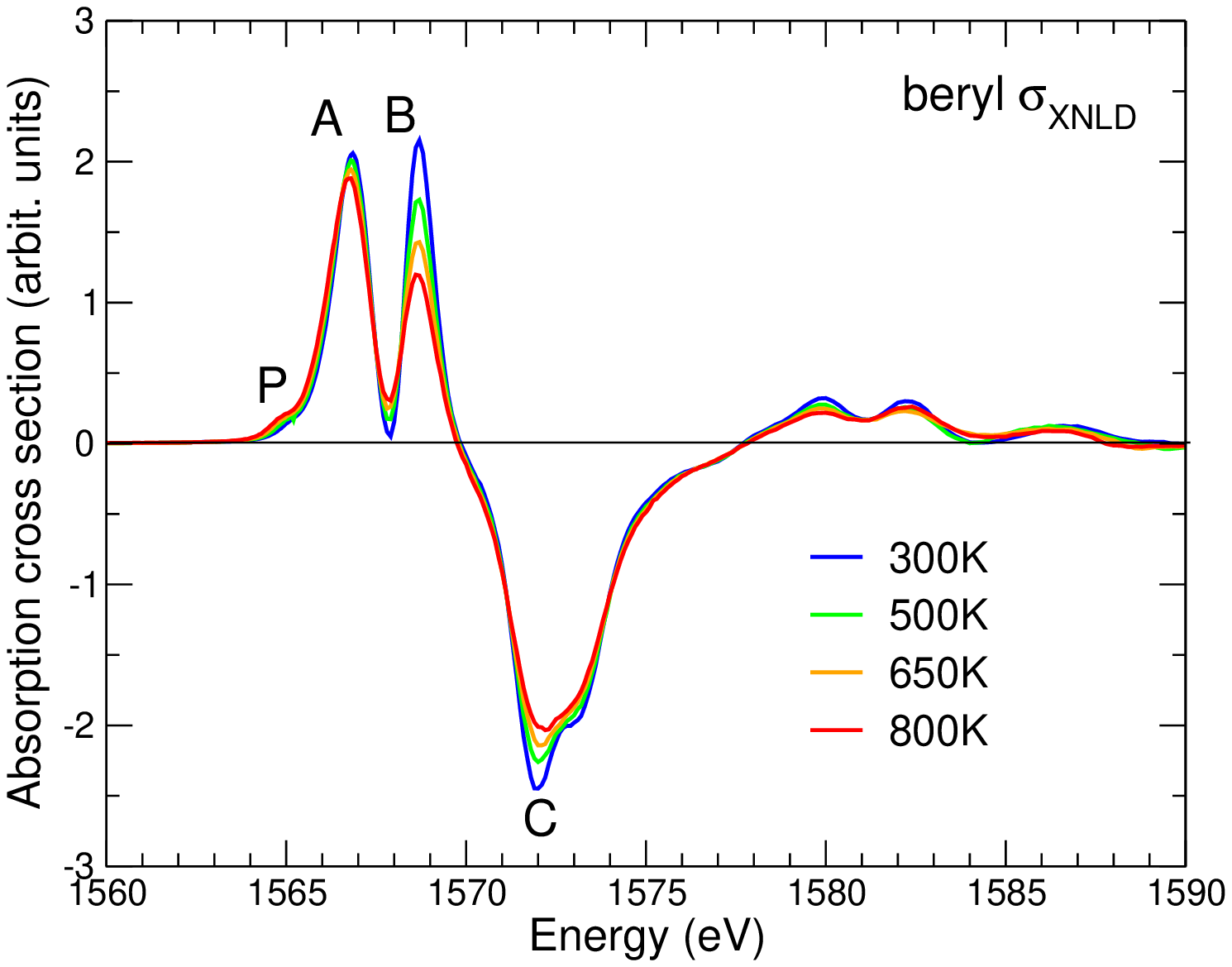}
\caption{\label{fig:xnld} Temperature dependence of the X-ray natural linear dichroism
(XNLD) measured at the Al $K$-edge in corundum (top panel) and beryl (bottom panel).
The cross section $\sigma_\mathrm{XNLD}$ is equal to 
the difference $\sigma_\parallel$-$\sigma_\perp$. Note that the $y$-axis range is twice larger
for beryl than for corundum.}
\end{figure}

The x-ray natural linear dichroisms (XNLD) for the various experimental temperatures 
are plotted in Fig.~\ref{fig:xnld}.
The XNLD signals of the two samples are large and their variations with temperature do
not change their general shape.
XNLD on the pre-edge is quite small so that temperature dependence can hardly be seen.
One mainly observes that some sharp XNLD maxima are reduced when the temperature increases:
A1 and A2 peaks for corundum and B peak for beryl.
One also notices that the rising edge and the maximum of the first XNLD feature is
shifted to lower energies. This effect is observed for both corundum and beryl 
and can also be detected on the parallel and perpendicular cross-sections. 
It is apparently a key feature of the temperature evolution of the spectra.
The features at high energies are only slightly modified.
Sum rules relates XNLD to the electric quadrupole distribution of the empty DOS
with $p$ symmetry on the aluminium site.\cite{Carra93}
The sum rules were first derived in the absence of electron-phonon
coupling and it is not yet clear whether the sum rules still hold when temperature is present.
A theoretical analysis of the temperature dependence
of XNLD sum rules is beyond the scope of the present paper but would eventually
give information on the modification of the electronic structure with the temperature.

\begin{figure*}[ht]
\centering
\includegraphics[width=0.40\textwidth]{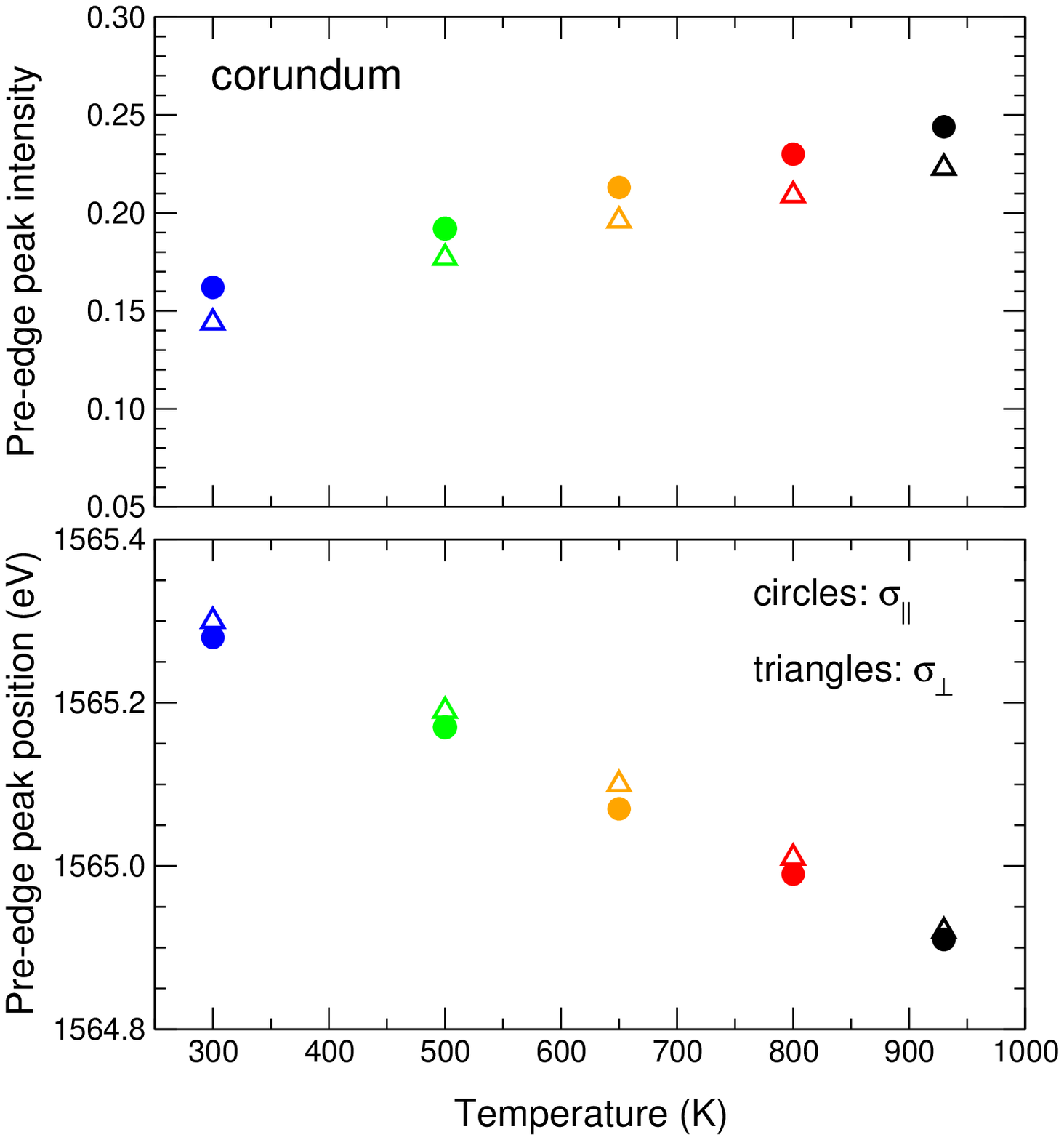}
\includegraphics[width=0.40\textwidth]{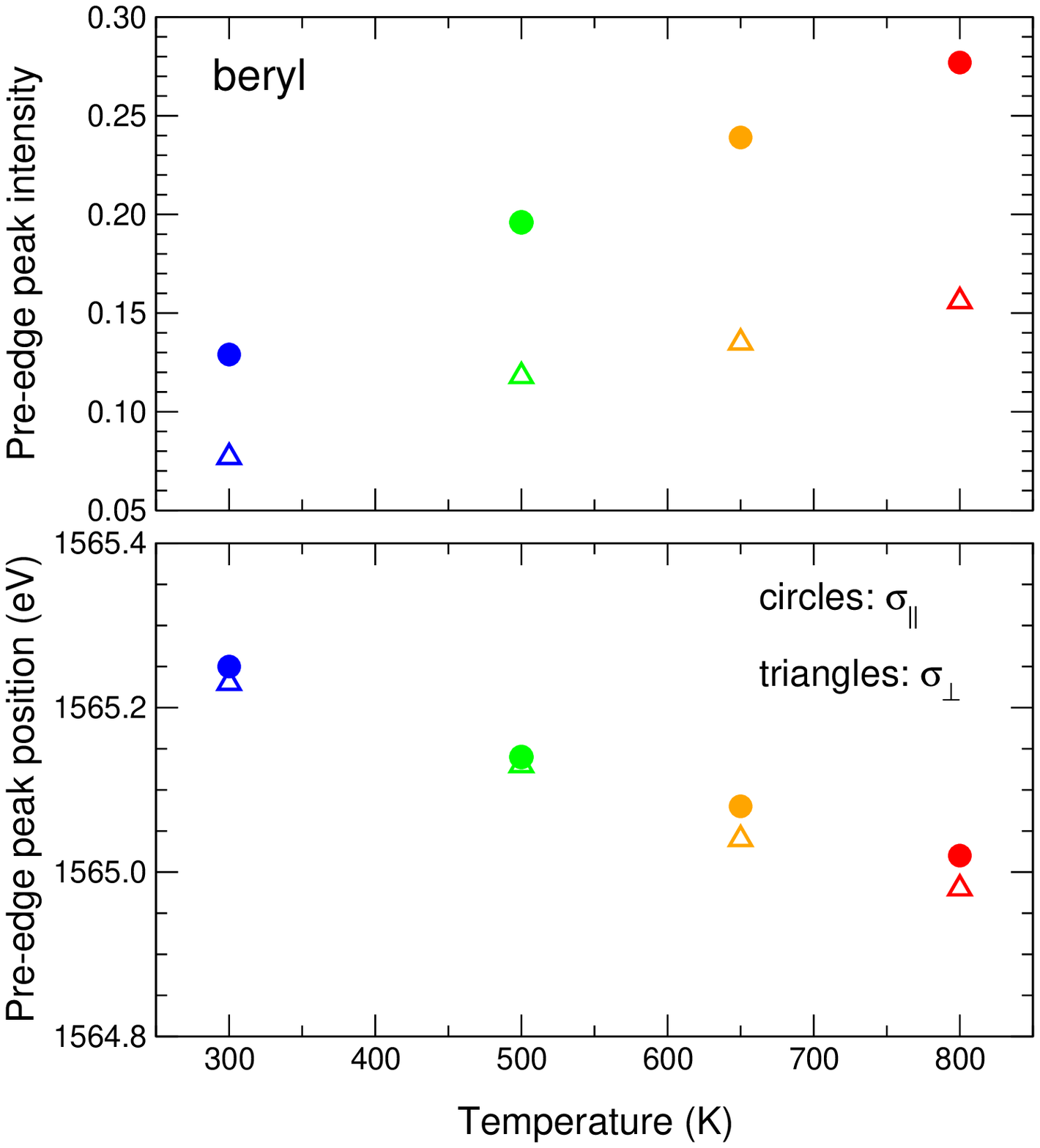}
\caption{\label{fig:pre-edge} Pre-edge intensity and energy position
for both orientations and for both samples as a function of temperature.
The $y$ axis units of the upper panels are consistent with those
of Fig.~\ref{fig:resexp}.}
\end{figure*}

In order to more quantitatively describe the behavior of the 
pre-edge peak, extraction was performed using Fityk software,\cite{fityk} 
which fits the main edge and pre-edge by the sum of an arctangent and a Gaussian. 
Figure \ref{fig:pre-edge} reports the fitted Gaussian intensities 
and center energies for each sample and each orientation.
These quantitative values are in good agreement with the foregoing 
observations, a global increase of intensity with temperature 
is noticeable. While both orientations 
seem to progress similarly in corundum, it is not the case in beryl 
where $\sigma_{\perp}$ intensities grow less rapidly than in $\sigma_{\parallel}$.
Concerning peak position, both compounds and orientations present a 
similar shift toward lower energies.

\section{Interpretation and discussion}
\label{sec:discussion}

In this section, theoretical results are presented and discussed in order 
to try to understand the effects of temperature observed
in the XANES spectra shown in Section~\ref{sec:resexp}.
First, a density of states (DOS) analysis is performed for both compounds
showing the involvement of the $3s$ empty states of the
Al absorbing atom in the process of pre-edge peak creation. 
Then, theoretical spectra obtained by using the three different methods 
described in section \ref{subsec:methodscalc} are shown and compared together.

\subsection{DOS calculations: the nature of the pre-edge peak}

\begin{figure*}
\centering
\includegraphics[width=0.40\textwidth]{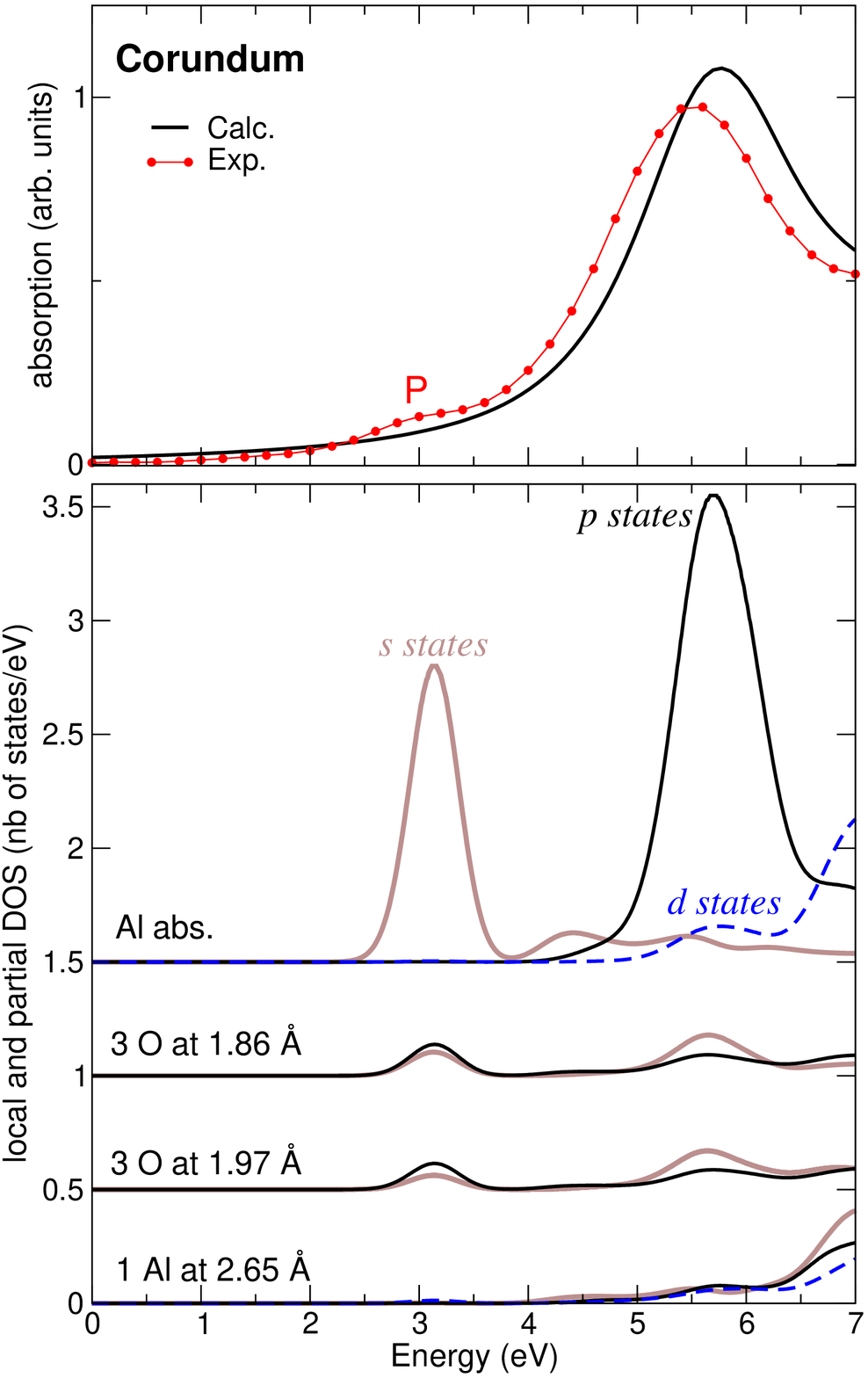}
\includegraphics[width=0.40\textwidth]{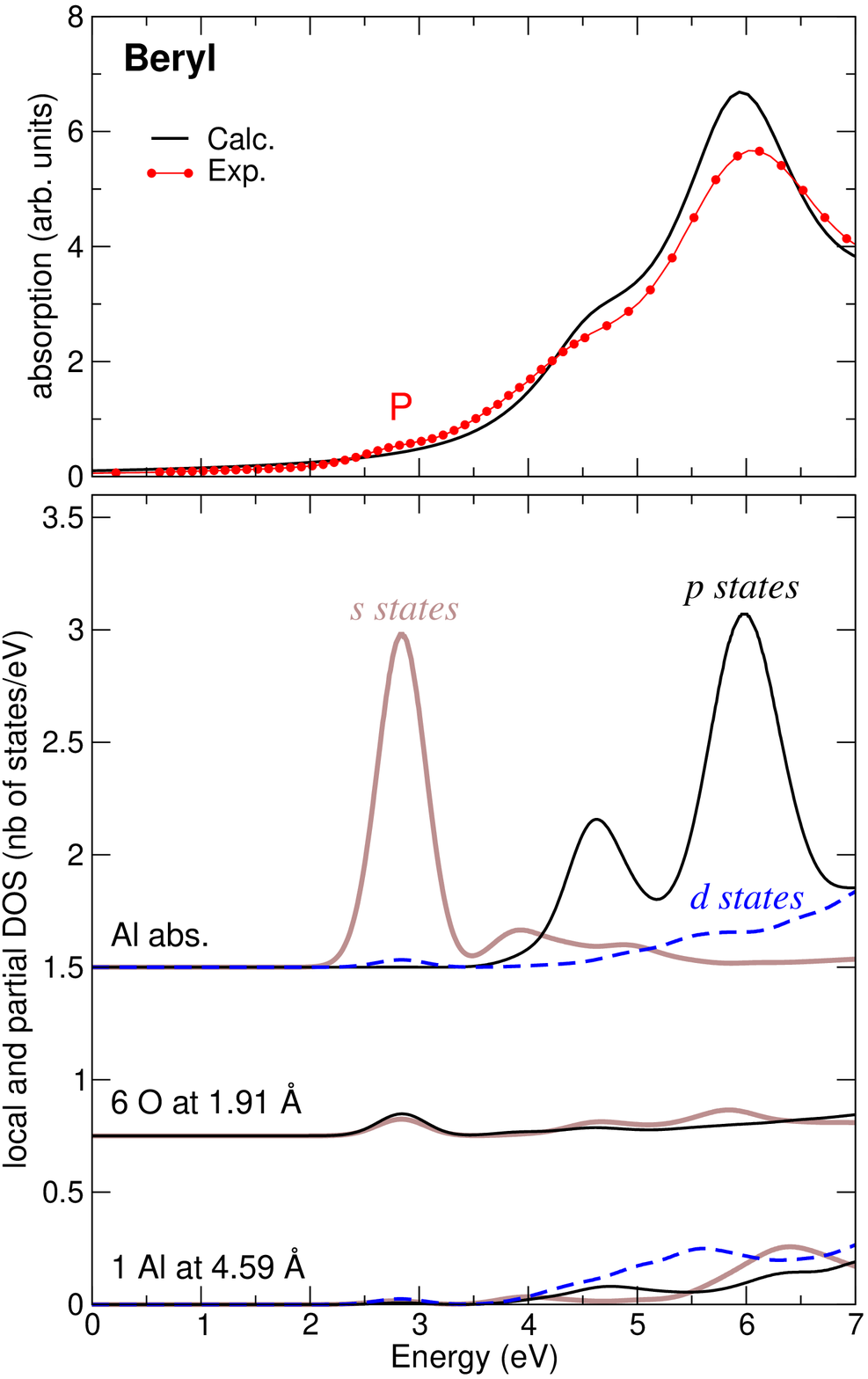}
\caption{\label{fig:DOS} Top: Calculated and experimental isotropic 
(2$\sigma_{\perp}$+$\sigma_{\parallel}$)/3 spectra at the Al $K$-edge 
of corundum (left) and beryl(right). Bottom: partial and local density 
of states of corundum (left) and beryl (right) calculated 
for the supercell including a 1$s$ core hole. 
The role of the Al 3$s$ of the absorbing atom is prominent in the pre-edge region.}
\end{figure*}

The pre-edge peak is not reproduced by purely electric 
dipole $1s$$\to$$p$ transitions, as shown in the top panels of Fig.~\ref{fig:DOS}.
Hence some transition toward non-$p$ states must be at stake.
The bottom panels of Fig.~\ref{fig:DOS} present partial 
and local density of states (DOS) of corundum and beryl. 
The $s$, $p$ and $d$ partial empty DOS are plotted for the absorbing 
aluminum atom (with a core hole), only the $s$ and $p$ for 
the first six oxygen neighbors, and again the $s$, $p$ and $d$ for
the aluminum next neighbor.
The similarity between the absorbing Al $3p$ local DOS and the XANES
spectra shows that the later is a good probe of the former.
In the pre-edge region, the role of the absorbing Al atom is clear 
as the pre-edge position coincides with its $3s$ projected DOS. 
In the case of beryl, a certain proportion of absorbing Al 3$d$ states
is also present in the pre-edge region. However, the 3$d$ 
states do not contribute to the absorption coefficient via electric 
quadrupole (1$s$$\rightarrow$3$d$) transitions. 
Indeed the calculated electric quadrupole transitions are
negligible in the whole XANES region, since they are found to be lower
than 0.25~10$^{-4}$ in the normalized units used in the top panels
of Fig.~\ref{fig:DOS}.
The $3s$ states of the absorber are hybridized with 
both $s$ and $p$ empty states of the oxygen neighbors 
while the aluminum next neighbors do not contribute to the 
pre-edge since their DOS are at higher energies.
The presence of the $1s$ core hole on the absorbing atom shifts 
the first empty states to the bottom of the conduction band and 
leads to the pre-edge and first peaks of XANES. 
Non-excited aluminum atoms do not contribute to the pre-edge 
feature of the spectrum.

\begin{figure*}
\centering
\includegraphics[width=0.49\textwidth]{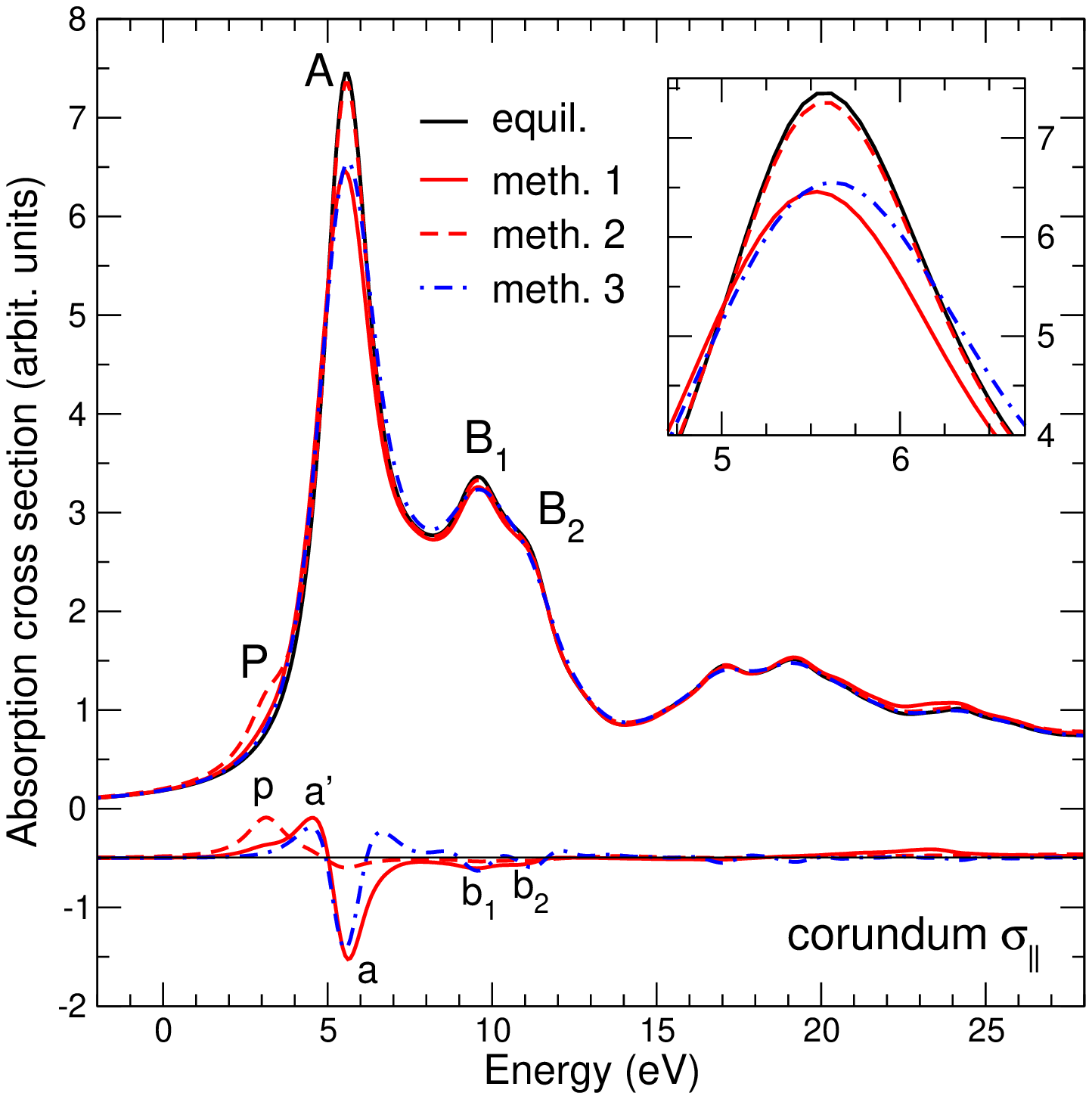}
\includegraphics[width=0.49\textwidth]{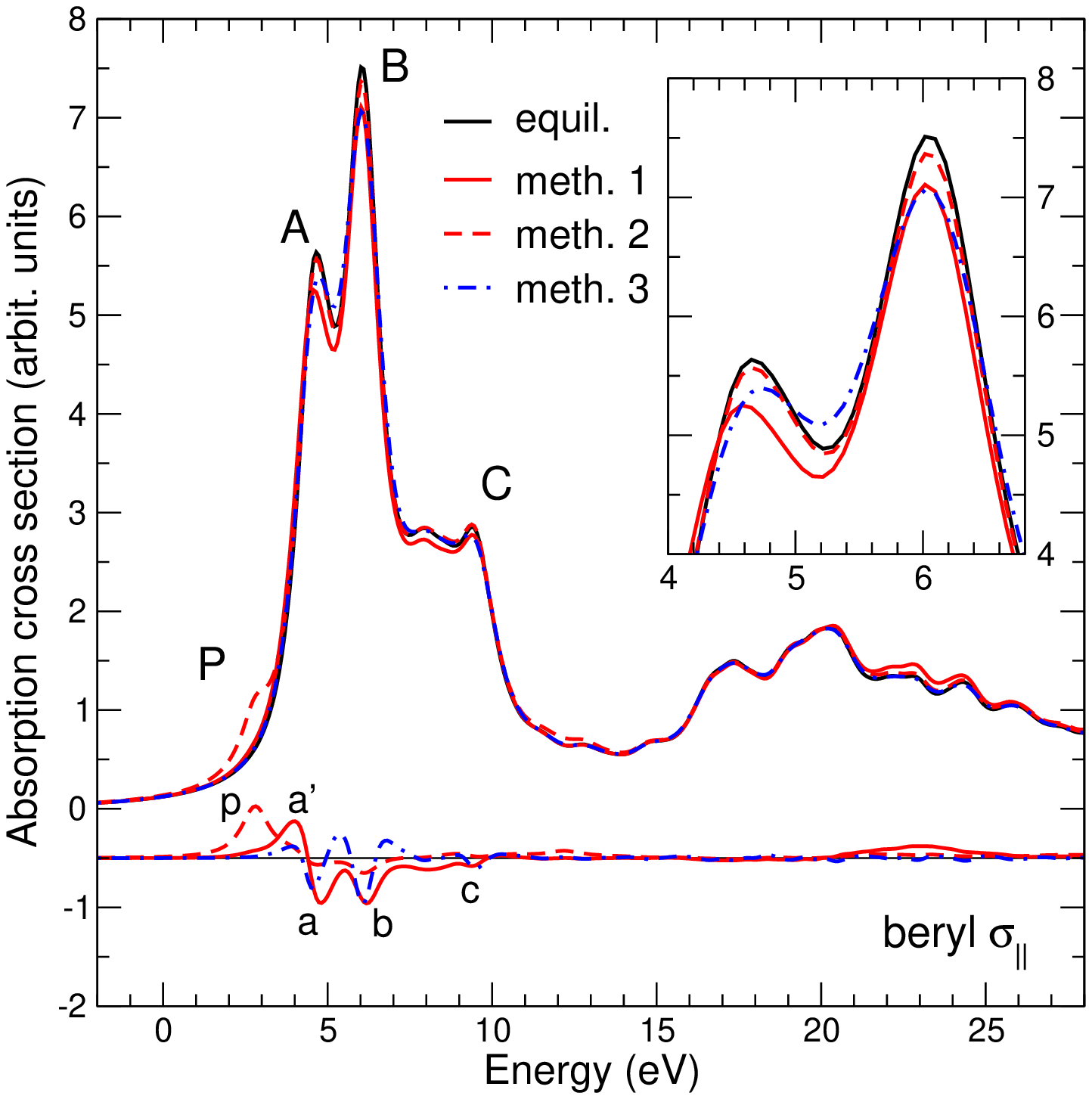}
\includegraphics[width=0.49\textwidth]{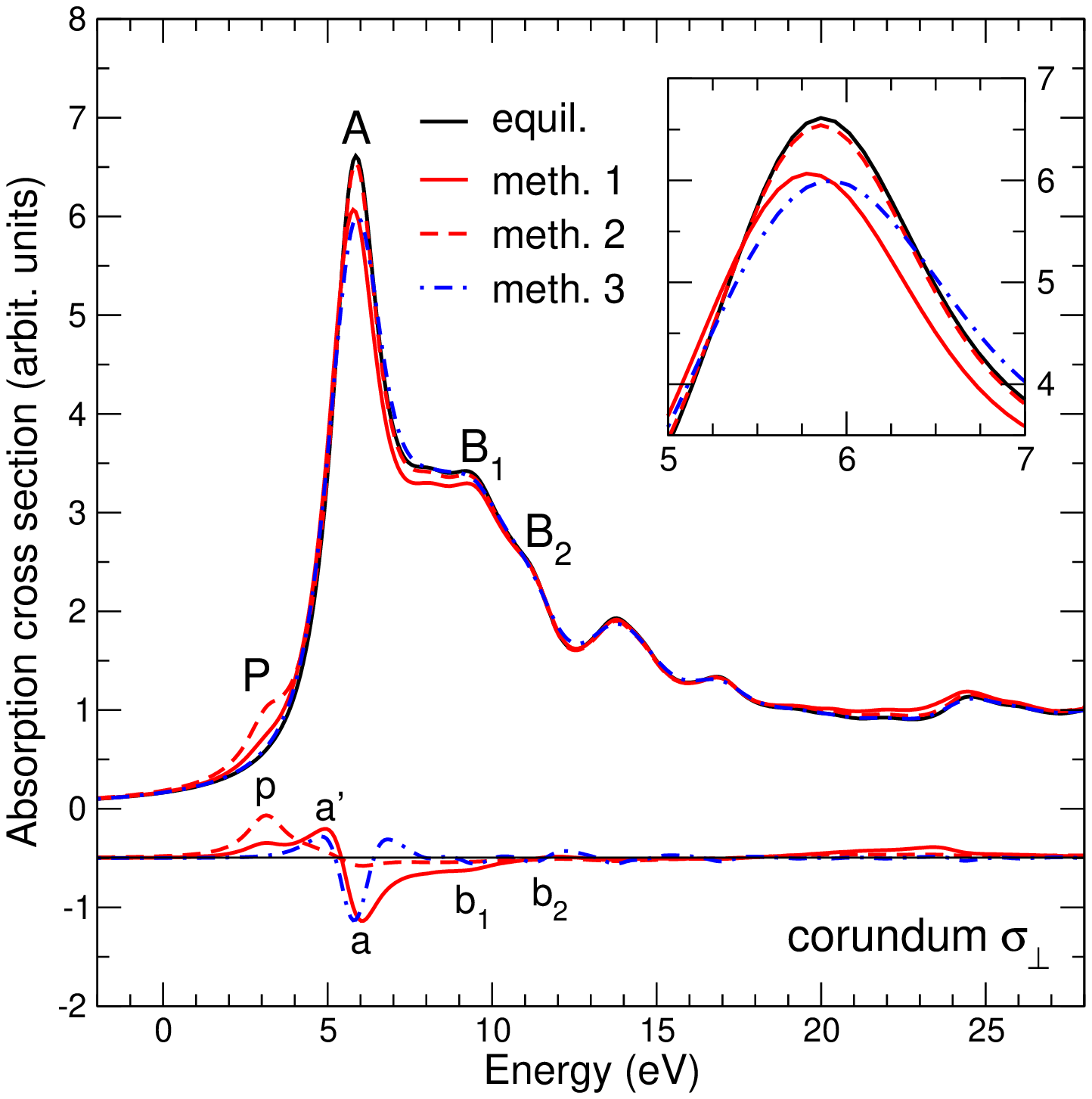}
\includegraphics[width=0.49\textwidth]{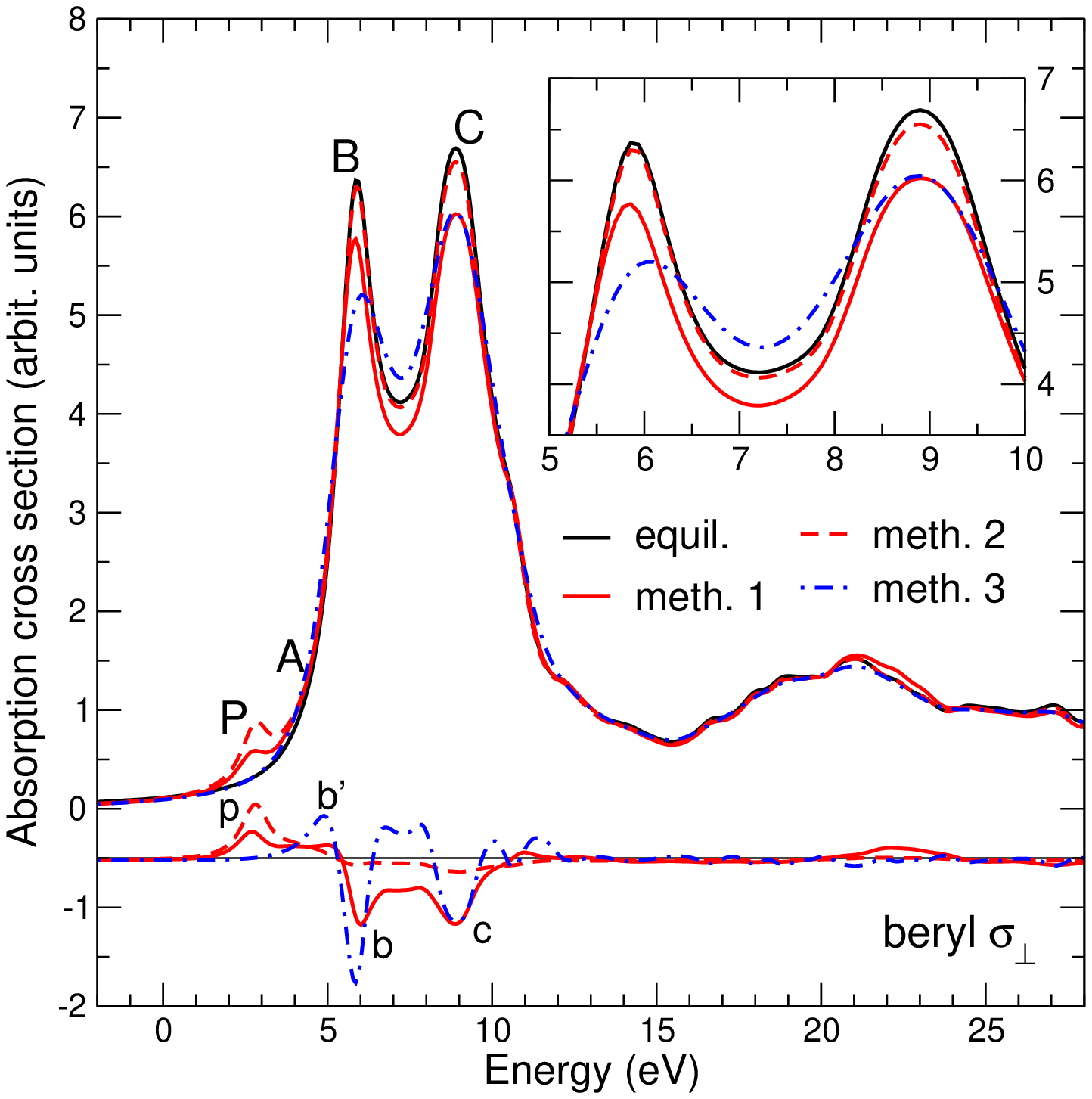}
\caption{\label{fig:calc_vib} Calculated $\sigma_{\parallel}$ and $\sigma_{\perp}$ 
spectra at the Al \textit{K} edge of corundum (left) and beryl (right) 
along with difference of each spectrum with respect to the equil. reference.
The `equil.' spectra correspond to DFT calculation performed with the atoms
at the equilibrium positions. 
The spectra labelled `meth. 1', `meth. 2' and `meth. 3' refer to the three methods of
vibration modeling 
described in section \ref{sec:methods}. 
The upper-right inserts use the same legend code 
and present a zoom on the main peak(s).}
\end{figure*}

\subsection{Theoretical XANES spectra: vibration modeling}

The DOS analysis indicates that
the modeling of the experimental Al $K$-edge XANES spectra has
to allow transitions toward the $3s$ states of the absorbing Al.
Both method~1 and method~2 presented in section~\ref{subsec:methodscalc} 
are able to achieve this goal. 
Figure~\ref{fig:calc_vib} displays the theoretical spectra obtained 
by the three methods described in section~\ref{subsec:methodscalc},
together with the spectra calculated with the atoms at their
equilibrium positions (labeled `equil.').
Difference spectra with respect to the `equil.'
reference are also plotted.
It should be noted that the `equil.' spectra do not strictly 
correspond to the 0K case since they do not take the zero point motion of 
nuclei into account. Experimental spectra measured at 20K
on the same samples (not shown here)
still exhibit the pre-edge peak, even without active 
phonon modes, and the pre-edge intensity at 20K is 
quite similar to that at 300K.
At room temperature nearly no phonon modes are active,\cite{Bialas75,Prencipe06} 
thus the pre-edge feature observable in the
20K-300K range seems to be essentially due to the quantum 
zero point motion effect.

Corundum $\sigma_{\perp}$ and $\sigma_{\parallel}$ `equil.' spectra
reproduce well the overall shape of the experimental spectra, except
in the pre-edge-region.
Switching on method~1 mostly affects the main edge \texttt{A} intensity
and shifts its position by 0.2~eV with respect to `equil.' spectra.
Peaks \texttt{a} and \texttt{a'} of the difference spectra are the
signature of these effects. Peak \texttt{B$_1$} intensity slightly
decreases for both orientations while peak \texttt{B$_2$} is nearly
unchanged. Small difference peaks \texttt{b$_1$} and \texttt{b$_2$}
confirm these points. A weak
pre-edge peak \texttt{P} seems to appear for both orientations.
Although method~1 is not able to grow a well-resolved pre-edge peak,
it provides difference spectra that are very similar to the 
temperature-dependent experimental difference spectra of Fig.~\ref{sec:resexp}
(left panels).
On the contrary, method~2 gives rise to a more important
pre-edge peak, which is in better agreement with experiment.
The difference peak \texttt{p} shows a really distinct Gaussian-like shape.
The rest of the spectrum is nearly unchanged and this
is confirmed by the flat difference spectra.
Method~3 spectra do not show a pre-edge but have a visible
impact on main edge peak \texttt{A}, by broadening it, lowering its intensity,
and shifting its energy to higher energies by 0.1~eV, i.e. in the opposite
direction as compared to temperature-dependent experiments.
These effects are visible on difference peaks \texttt{a} and \texttt{a'}.

Beryl $\sigma_{\parallel}$ and $\sigma_{\perp}$ `equil.' spectra
also reproduce the experimental spectra quite well, except
for some of the peak intensities. While peaks \texttt{A} and
\texttt{C} are of similar amplitude in $\sigma_{\parallel}$
experimental spectrum, the intensity of peak \texttt{A} is twice
the one of peak \texttt{C} in the `equil.' spectrum. On the opposite,
peaks \texttt{B} and \texttt{C} have similar intensity in 
$\sigma_{\perp}$ `equil.' spectrum while peak \texttt{B} is 1.5 times
less intense than peak \texttt{C} in the 300K experimental spectrum.
By using method~1 for $\sigma_\parallel$, no pre-edge peak 
seems to arise, peak \texttt{A} is decreased and shifted to lower energy
by 0.2~eV, peak \texttt{B} is also decreased and peak \texttt{C} 
remains stable. These effects are also visible through the shape of the
difference spectrum. Thus vibrations treated within method~1 are not able 
to well reproduce the temperature-dependence
trend observed in the top-right panel of Fig.~\ref{fig:resexp}.
In the case of $\sigma_\perp$, method~1 leads to theoretical spectrum 
in better agreement 
with the x-ray absorption temperature-dependence seen in the bottom-right
panel of Fig.~\ref{fig:resexp}. Indeed, method~1 enables a well resolved 
pre-edge peak \texttt{P} to rise (clearly confirmed by difference peak \texttt{p}),
and a very weak peak \texttt{A} to appear. It also moves \texttt{B} to lower energy
(0.2~eV) and decreases the intensity of peaks \texttt{B} and \texttt{C}.
Here again, method~2 has no influence on peak position,
it nearly does not change the amplitude of peaks \texttt{A},
\texttt{B}, \texttt{C} in $\sigma_{\parallel}$ and \texttt{B}, \texttt{C}
in $\sigma_{\perp}$. However it creates the pre-edge \texttt{P}
at the right energy, but with too high an intensity.
The value of the $U$ parameter used in method~2 might be overestimated
for beryl and yielded a too pronounced pre-edge peak.
Method~3 decreases the amplitude of main peaks. It also shifts peak \texttt{A}
of $\sigma_{\parallel}$ and \texttt{B} of $\sigma_{\perp}$ towards
higher energies that, as in corundum, is the opposite of what is seen experimentally.
A slight broadening appears and is more visible in $\sigma_{\perp}$.

To summarize,  method~1 leads to difference spectra that reproduce quite well
the temperature-dependent trend observed experimentally in most cases.
Nevertheless the impact of vibrations as treated by method~1 
is not sufficient to systematically reproduce the experimental pre-edge feature. 
The $p$-$s$ hybridization induced by this method is not satisfactory enough.
In method~1, the vibration modeling results in integration, over a cubic volume
around the absorbing atom, of cross sections
calculated for absorbing Al isotropically displaced inside this volume.
In other words, this method equally considers
the contributions of Al motion along all directions.
Some Al displacements would be favored by taking into account the phonon
modes of the material, what could substantially improve the agreement between
theoretical and experimental spectra at least in the pre-edge region.
Method 2 spectra all exhibit a well-defined 
pre-edge peak, corresponding to dipole 1$s$-3$s$ transitions, 
which are allowed by the displacement 
of the $1s$ wave-function.\cite{Brouder10,Cabaret09}
Nevertheless, method~2 has nearly no impact on the peak intensity 
above the edge, and is unable to produce any shift of the features.
This last result was expected because of the use of the
crude Born-Oppenheimer (BO) in method~2. Within the crude BO, the transition
energies are evaluated for the atoms at their equilibrium positions, hence
the spectral features remain at the same energy positions as in 
`equil.' spectra.
Method~3 spectra, as expected, do not exhibit any pre-edge peak, 
and present main-edge peaks less intense and shifted in energy.
However the energy shift goes the wrong way, i.e. to higher energies. 
Therefore, a simple convolution of the `equil.' spectra seems
to be inappropriate to account for the temperature-dependence of 
the XANES spectra observed experimentally. 

Thermal expansion effects have been evaluated
in the case of corundum by performing a XANES calculation for the 
structural parameters refined at 2170K 
(153K below the melting point).\cite{Ishizawa80}
A contraction of the highest energy XANES features has been noticed.
Such an effect was expected according to the predictions of 
the Natoli's rule.\cite{Natoli83}
However, it is contradictory with what has
been experimentally observed when temperature is increased (Fig.~\ref{fig:resexp}). 
Therefore, the thermal expansion is unable to explain 
the temperature-dependence of the Al $K$-edge XANES spectra in 
corundum and beryl.

The calculations carried out in this study clearly show the crucial role of vibrations
in the pre-edge region and partially explain the temperature-dependence observed in the
whole XANES region. However, none of the three methods used here 
is fully satisfactory. A great improvement would be to use PIMD
as Schwartz \textit{et al.} did in the case of two isolated organic molecules.\cite{Schwartz09}
But the computational cost of PIMD simulations in solids, such as corundum
or beryl, would be a serious limiting factor.
A more reasonable way to account for the thermal fluctuation of XANES 
at the Al $K$-edge would consist of the generation of
atomic configurations of the whole crystal
from the dynamical matrix of the system.
The temperature-dependent XANES theoretical spectrum would 
then result in an average of cross-sections calculated for a large number of configurations. 
An analogous methodology has been successfully employed to account for 
the temperature-dependence of nuclear magnetic resonance chemical shift in MgO.\cite{Rossano05}

\section{Conclusion}
\label{sec:conclusion}
In this study, temperature-dependent Al $K$-edge XANES spectra
of corundum and beryl
have been presented for temperature ranging from 300K to 930K
for the first time.
The XANES spectra were measured on single-crystals with the polarization
vector of the x-ray beam parallel and perpendicular to the high symmetry
axis of the minerals, allowing
the investigation of the temperature dependence of the $\sigma_\parallel$
and $\sigma_\perp$ components.
This series of experiments shows that the pre-edge peak is very sensitive to
thermal fluctuations: the pre-edge peak intensity grows and its position
shifts to lower energy as temperature increases. 
These variations do not depend much on the x-ray polarization.
Thermal fluctuations are also visible above the pre-edge region, through
an intensity decrease of the main features and, in the case of corundum, 
through a slight shift to lower energy of the first main peak.

First-principles DFT calculations have confirmed the Al 3$s$ nature of the
pre-edge peak in both minerals, and have shown that the introduction
of vibrations within the XANES calculation gives rise to a pre-edge peak.
Our experimental and theoretical results on the Al $K$ pre-edge
in corundum and beryl suggest that vibrations (the zero point motion 
and also temperature)
could be able to induce a pre-edge feature  
at the Al $K$-edge in other minerals. This conclusion might bring 
a reinterpretation of the appearance of a pre-edge peak 
observed at the Al $K$-edge in zeolites when temperature
is increased, which was interpreted as the signature 
of three-fold coordinated Al.\cite{vanBokhoven03,vanBokhoven05,Agostini10}

We have used three existing theoretical methods to take nuclear motion
into account. These methods have provided useful information to 
understand the origin of the spectral modifications observed
in the temperature-dependent experimental XANES spectra. However, they were
found to be inappropriate to fully account for the thermal fluctuation
of XANES.
This knowledge will be particularly useful for upcoming works,
giving insight of what has to be considered in the modeling of 
temperature-dependent XANES spectra.

\section{Acknowledgements}
We are grateful to Steve Collins for showing us the temperature
dependence of Ti $K$-edge spectra that he measured on rutile at beamline I16 at Diamond Light Source.
This work was granted access to the HPC resources of IDRIS 
under the allocation 2011-1202 and 2012-100172 made by GENCI 
(Grand Equipement National de Calcul Intensif). 
Experiments were performed on the LUCIA beamline at SOLEIL Synchrotron, 
France (proposal numbers 20100888 and 99110023). We are grateful to the SOLEIL staff 
for smoothly running the facility.
We also acknowledge Pierre Lagarde for fruitful discussions.

\end{document}